# The Effect of Hole Transporting Layer in Charge Accumulation Properties of p-i-n Perovskite Solar Cells

Fedros Galatopoulos, Achilleas Savva, Ioannis T. Papadas, Stelios A. Choulis[1]

Molecular Electronics and Photonics Research Unit, Department of Mechanical Engineering and Materials Science and Engineering, Cyprus University of Technology, Limassol, 3603 (Cyprus).

The charge accumulation properties of p-i-n perovskite solar cells were investigated using three representative organic and inorganic hole transporting layer (HTLs): a) Poly(3,4-ethylenedioxythiophene)-poly(styrenesulfonate) (PEDOT:PSS, Al 4083), b) copper-doped nickel oxide (Cu:NiOx) and c) Copper oxide (CuO). Through impedance spectroscopy analysis and modelling it is shown that charge accumulation is decreased in the HTL/Perovskite interface, between PEDOT:PSS to Cu:NiOx and CuO respectively. This was indicative from the decrease in double layer capacitance (Cdl) and interfacial charge accumulation capacitance (Cel), resulting in an increase to recombination resistance (Rrec), thus decreased charge recombination events between the three HTLs. Through AFM measurements it is also shown that the reduced recombination events (followed by the increase in Rrec) is also a result of increased grain size between the three HTLs, thus reduction in the grain boundaries area. These charge accumulation properties of the three HTLs have resulted in an increase to the power conversion efficiency between the PEDOT:PSS (8.44%), Cu:NiOx (11.45%) and CuO (15.3%)-based devices.

---

[1] stelios.choulis@cut.ac.cy

Over the last couple of years perovskite-based solar cells have shown advancements, with an incredibly fast power conversion efficiency (PCE) improvement from 3.8% to 21%.[1,2] The high PCE values are a result of several attractive features that hybrid perovskite semiconductors offer, even if casted from solution at low processing temperatures.[3] Their main solar cell related attractive features are the low exciton binding energy,[4,5] tunable bandgap,[6,7] high absorption coefficient,[8] long carrier diffusion length[9] and high carrier mobility.[10]

The p-i-n structured perovskite solar cell is distinguished compared to other perovskite solar cell device architectures due to its facile processing. In this structure, the perovskite active layer is placed between two carrier selective electrodes and is in direct contact with the HTL of the transparent bottom electrode and the electron-transport layer (ETL) of the opaque back electrode. Each electrode is responsible for allowing only a specific type of carrier to be collected by blocking the opposite carrier type.[8] Efficient carrier generation, transport and collection are crucial for proper photovoltaic operation.

One of the most important aspects of the solar cell device performance is the crystallinity of the perovskite active layer. A critical factor affecting the crystallinity of the perovskite sensitizer is the layer on which the perovskite photoactive layer is grown from solution[11]. It has been shown that the under-layer has a major influence on the perovskite crystal orientation, grain size and morphological defects. These factors are crucial for highly efficient perovskite solar cells.[11,12]. Large perovskite grain size is generally believed to result in increased carrier diffusion length and mobility of carriers.[12] Furthermore, a crystal structure with large grains can provide good surface coverage and small grain boundaries area, which are also desirable since they can minimize the presence of pinholes and charge traps, effectively reducing carrier recombination.

Incorporating HTLs in perovskite solar cells has been a crucial part towards the achievement of high performance devices. The main features of an efficient HTL are: a) efficient electron blocking, b) efficient hole extraction. These features can have a major impact to device parameters, e.g. Voc via quasi Fermi level splitting.[13] Different types of materials have been studied and used as HTLs thus far, including inorganic, polymeric[14] and small organic molecules.[14,15] Considerable effort has been put in the process of HTL optimization in terms of stability and efficient semiconducting properties. An example is the use of cobalt based dopants to enhance the charge carrier mobility of the widely used spiro-OMeTAD[15].

The focus of this study is to analyze the effect of the HTL in the charge accumulation of a p-i-n structure perovskite solar cell. The exact transport and charge collection mechanisms in perovskite solar cells are not yet completely understood. Impedance spectroscopy has been used to provide valuable information about the physical processes manifesting inside the perovskite-based solar cells.[8] By studying these processes, both from various experimental methods and by fitting the data in an equivalent circuit model (ECM) we were able to extract valuable information regarding the charge accumulation at the p-i-n structure perovskite solar cell device.

In the present study, three p-i-n $CH_3NH_3PbI_3$–based devices with different HTLs were fabricated and extensively analyzed using a series of measurements. For all the devices the ETL used was [6,6]-Phenyl-C70-butyric acid methyl ester (PC[70]BM). A thick (~50 nm) Al-doped zinc oxide (AZO) interlayer was incorporated between PC[70]BM and Al. An AZO interlayer has shown to increase the electron selectivity of p-i-n planar perovskite-based solar cells.[6]

The three devices under study have the structure: a) ITO/PEDOT:PSS/$CH_3NH_3PbI_3$/PC(70)BM/AZO/Al (PEDOT:PSS-based) b)

ITO/Cu:NiOx/CH$_3$NH$_3$PbI$_3$/PC(70)BM/AZO/Al (Cu:NiOx-based) c) ITO/CuO/CH$_3$NH$_3$PbI$_3$/PC(70)BM /AZO /Al (CuO-based) as shown in Fig. 1a.

**FIG.1. a)** Graphical presentation of the solar cell layers, **b)** AFM measurements of ITO/PEDOT: PSS/CH$_3$NH$_3$PbI$_3$, **c)** ITO/Cu:NiOx/CH$_3$NH$_3$PbI$_3$ and **d)** ITO/CuO/CH$_3$NH$_3$PbI$_3$.

The perovskite active layer was fabricated following the methodology reported by Liang et.al.[16] with an estimated active layer thickness of ~200 nm for all the devices. The thickness measurements were performed using a Veeco Tektak 150 profilometer. Atomic force microscopy (AFM) measurements were performed using Easy Scan 2-Nanosurf in tapping (phase contrast) mode with 5x5μm image size. From Figure 1b, c and d, it is observed that CH$_3$NH$_3$PbI$_3$ perovskite photoactive layer exhibits significantly different grain size when fabricated on top of different HTLs. The CH$_3$NH$_3$PbI$_3$ grain size is estimated from the AFM measurements at ~625nm, ~450nm and ~180nm when fabricated on top of CuO, Cu:NiOx and PEDOT:PSS, respectively. As previously mentioned the increased grain size, and therefore reduced grain boundaries area of the perovskite photoactive layers, influences the charge transport properties and subsequently the performance of corresponding devices.[12] The exact effect of the grain size and grain boundaries towards charge accumulation is discussed further down this manuscript.

Representative devices from each structure were chosen and characterized using J-V and impedance spectroscopy. The devices with the highest PCE values were chosen for study from each architecture. The active area for all the devices was at 0.09 cm$^2$. No hysteresis was observed, between the forward and reverse scan of the current density-voltage (J/V) characteristics, for all the devices under study. The preparation of PEDOT: PSS and Cu:NiOx-based devices was based on the methodology of Jung et. al.[17] resulting in a thickness of ~40 nm for each layer. The CuO-based synthesis and fabrication details can be found in our recent study.[18] The thickness of the CuO interfacial layer was measured at 15 nm. J/V measurements were performed using a solar simulator from Newport equipped with a Xenon lamp at 100 mW/cm$^2$ providing an AM 1.5 G spectrum and a scan rate of 0.1V/s. Impedance measurements were performed using a Metrohm Autolab PGSTAT 302N. A red LED (at 625 nm) was used as the light source. A small AC perturbation of 20 mV was applied to the devices and the different current output was measured throughout a frequency range of 1MHz-1Hz. The steady state DC bias was kept at 0 V throughout the impedance spectroscopy experiments.

**FIG.2. a)** Current density versus voltage (J-V) characteristics under illumination, **b)** Nyquist plots **c)** Equivalent circuit model and simulation curve used for analyzing PEDOT: PSS based devices, **d)** Equivalent circuit model and simulation curve used for analyzing Cu:NiOx and CuO based devices.

Fig. 2a shows the J-V characteristics of the three devices under study. The plots show a clear difference in the Voc and Jsc of the three p-i-n solar cells under comparison. Although devices with the PEDOT:PSS have better reproducibility, the solar cells exhibit relatively low efficiency with values of Voc=0.85 V, Jsc=13.19 mA/cm$^2$, FF=75.6 % and PCE=8.44 %. In contrast, Cu:NiOx-based devices exhibit increased Voc=1.03V, Jsc=14.78mA/cm$^2$, FF=75% and PCE=11.45%. Finally, CuO-based devices exhibit Voc=1.09 V, Jsc=18.2 mA/cm$^2$, FF=77% and PCE=15.3%. From the values shown in Table I, the best performing device in terms of PCE was the CuO-based device followed by Cu:NiOx-based device and then the PEDOT:PSS-based device. The origin of the PCE improvement between the three p-i-n solar cells could be attributed in two main factors: a) The lower work function of CuO (5.6 eV)[19] and Cu:NiOx (5.3 eV)[20] compared to PEDOT:PSS (4.9 eV)[21], offering a more favorable energy level alignment with the perovskite (Ec= 3.9 eV, Ev=5.4 eV)[22] and thus more efficient hole extraction, leading to increased Voc and Jsc for the corresponding solar cells, b) the increased grain size of the $CH_3NH_3PbI_3$ when fabricated on Cu:NiOx and CuO-based devices compared to PEDOT:PSS-based devices (see Fig.1). The increased grain size (and thus the reduced grain boundaries area) potentially leads to decreased carrier recombination due to decreased charge trap densities as reported elsewhere.[12,23] The photovoltaic parameters as well as fitted parameters from impedance spectroscopy are shown in Table I.

**TABLE I**. Extracted solar cell parameters from the characterization and simulation of the device.

| Device HTL | Voc (V) | Jsc (mA/cm²) | FF (%) | PCE (%) | Rtr (Ω) | Rrec (Ω) | Cdl (F/cm²) |
|---|---|---|---|---|---|---|---|
| PEDOT:PSS | 0.85 | 13.19 | 75.6 | 8.44 | 534 | 9000 | $2.8 \times 10^{-5}$ |
| Cu:NiOx | 1.03 | 14.78 | 75 | 11.45 | 467 | 11000 | $1.5 \times 10^{-5}$ |
| CuO | 1.09 | 18.2 | 77 | 15.3 | 2350 | 25000 | $3.5 \times 10^{-7}$ |

To further understand the effect of the HTL in charge accumulation of p-i-n perovskite solar cells, impedance spectroscopy has been used. Fig. 2b shows three representative Nyquist plots of the devices as well as their simulated fits from the ECMs. It can be clearly seen that the plots follow the general shape of two frequency responses for perovskite based solar cells.[24,25] The feature at high frequencies has been previously attributed to the charge transport resistance (Rtr) of the ETL and HTL as well as their interfaces with the perovskite absorber[24,25,26] whereas the feature at low frequencies has been previously attributed to the recombination resistance (Rrec) and ionic diffusion.[24,25] Interestingly, the three p-i-n solar cells incorporating different HTLs exhibit different impedance responses at the intermediate frequency range. As it can be seen from the Nyquist plots the PEDOT: PSS-based device shows an inductive loop manifesting at intermediate frequencies. An inductive loop in an impedance response often indicates sudden change in the sign of capacitance, triggered by several phenomena (e.g. charge depopulation), manifesting as negative capacitance.[27] This inductive loop is a feature previously observed in battery systems attributed to the constant formation of the solid electrolyte interphase products, which involves ionic movement[28] and has also been previously observed in n-i-p perovskite solar cell at intermediate frequencies[26]. The absence of this feature was previously believed to be a sign of charge accumulation at the device interfaces,[26] however in this study we have observed this feature only in organic PEDOT:PSS-based HTL devices and it is absent from the more efficient devices based on inorganic Cu:NiOx and CuO HTLs. It has been reported that perovskite solar cells are mixed ionic and electronic conductors and one of the main route for the mobile ions movement is through the grain boundaries of the perovskite photoactive layers.[29] Since the devices with Cu:NiOx and CuO exhibit reduced grain boundaries area, a possible explanation can be that there are less pathways for ion migration in the device. It is beyond the scope of this paper to study the exact origin of the inductive loop. However, the absence of this feature from our best performing Cu:NiOx and CuO devices leads us to believe that a potential explanation could lie in the ionic nature of perovskite solar cells using inorganic HTLs.

Because of the different shape of the Nquist plots we were unable to use a universal ECM for the three devices under investigation. Instead, two different ECMs have been used, which are shown in Fig. 2c and 2d. The blueprint for the ECM used was based on the circuit model described from Bag et.al.[21] Fig. 2c presents the circuit that was used to model PEDOT:PSS-based devices, whereas Fig. 2d shows the circuit that was used to model Cu:NiOx and CuO-based devices. Both models share a variety of similar elements, however they also have distinct differences. Rs denotes the series resistance from external circuitry (from wires and equipment). Rtr and Rrec are attributed to the charge transport resistance and recombination resistance respectively. These elements are connected in parallel with Cg, which is the geometrical capacitance arising from the bulk perovskite and are used to model the high frequency component of the Nyquist plots. The parallel connection of Rct, Ws and Cdl was used to model the low frequency component of the Nyquist plots. Rct represents the charge transfer resistance whereas Ws is the Warburg diffusion element used to model ionic diffusion in the device. Cdl is the double layer capacitance arising from charge accumulation at the interfaces (from both ionic and electronic components). In order to effectively model the inductive loop observed in the PEDOT:PSS-based device, an R-L circuit was used in series with Rtr (Fig. 2c). The circuit in Fig. 2d was used to model the low frequency component in the Nyquist plots for Cu:NiOx and CuO-based devices. This circuit contains the elements that were initially discussed with the exception of replacing Cdl with a constant phase element (CPE). The simulated values of the key features of the equivalent models (ECM) that were used are summarized in Table I. From Table I it can be seen that the transport resistance (Rtr) has a comparable value for the devices with Cu:NiOx-based device (467 Ω) and PEDOT:PSS-based device (534 Ω), whereas CuO-based device yields a much larger transport resistance (Rtr) size (2350 Ω). Since the only variable changing between the devices is the HTL used, from the previous observations and Rtr values we can conclude that the conductivity of CuO is poorer compared to the other two HTLs, whereas the conductivity of PEDOT:PSS and Cu:NiOx is comparable. This is in agreement with the values from previous studies that report the following conductivity values: PEDOT:PSS Al 4083~0.9x10$^{-3}$ S/cm[30], Cu:NiOx~1.25x10$^{-3}$ S/cm[17] and CuO~10$^{-6}$ S/cm[31]. From Fig.2b as well as the value extracted from the ECM, an increase to recombination resistance (Rrec) for Cu:NiOx-based (11000 Ω) and CuO-based (25000 Ω) devices compared with PEDOT:PSS-based device (9000 Ω) was observed. The increase in Rrec is directly related to our previous observations. The progressively increased grain size and reduced grain boundaries area between the three different under layers indicate that less charges are accumulating in the grain boundaries of the devices, resulting in decreased carrier recombination. Furthermore, due to the more favorable energy level alignments between the three HTLs, previously discussed, the extraction of holes is energetically favored and thus decreased interfacial charge accumulation occurs, resulting in a decrease in carrier recombination in the HTL/Pvsk interface. This is in accordance with the increase in Rrec between the three devices, which denotes less frequent recombination events.

Furthermore, we observe that the double layer capacitance (Cdl) is decreased for CuO-based device (3.5x10$^{-7}$ F/cm$^2$) compared with Cu:NiOx-based device (1.5 x10$^{-5}$ F/cm$^2$) and PEDOT:PSS-based device (2.8x10$^{-5}$ F/cm$^2$). Cdl arises from charge accumulation at the interfaces from both ionic and electronic accumulation, although under light, Cdl is predominantly affected by the accumulation of photo-generated carriers.[26] Since Cdl is tied to charge accumulation at the interfaces, having a small value is favorable for increased device performance. The reduction in the value of Cdl that we observe between the three devices is directly correlated to our previous observations regarding the decrease in charge accumulation at the HTL/Pvsk interface. Both observations regarding the decreased charge accumulation at the grain boundaries area as well as the HTL/Pvsk interface have a direct impact to the increased Voc and Jsc of the corresponding solar cell devices.

Apart from the above observations, to be able to obtain good fits for the CuO based device the Cdl had to be changed from an ideal capacitor to a constant phase element (CPE). CPEs are usually used to model the behavior of a non-ideal capacitor. This could mean that the capacitance of a layer or interface is not constant and varies between regions of the layer. This inhomogeneous nature has a plethora of explanations such as layer roughness[32], non-uniform current distribution[33], variations in thickness[34] e.t.c. In this particular case, we assume that this inhomogeneous nature arises from non-uniform charge distribution resulting from the nature of the very thin CuO nanoparticulate HTL. In the case of CuO, the nanoparticles (NPs) were synthesized by solvothermal method[18] and monodispersed NPs with size of 5-10 nm were formed. The variations in NP size and film compactness, as well as the low intrinsic electrical conductivity of CuO, could result in inhomogeneous hole collection at the perovskite/CuO interface and thus inhomogeneous capacitance distribution within the device area. The latter can also be the origin of the increased Rtr observed for the CuO-based devices. The Cu:NiOx was synthesized following a combustion method. This method requires the reaction of inorganic salts as oxidizer with an organic fuel.[17] Despite the higher conductivity values of Cu:NiOx compared to CuO, this method can leave residual unreacted organic compounds in the final metal oxide, which would contribute in the non-uniform charge distribution. This parameter could explain the need to use a CPE in doped inorganic HTLs. In contrast perovskite solar cells using the high conductive and relative thick organic PEDOT:PSS[35] buffer layers have a relatively uniform charge distribution and homogeneous charge collection. Thus a homogeneous capacitor in the ECM can provide a good fit.

To better understand the charge accumulation mechanisms in p-i-n perovskite solar cells capacitance-frequency (C-F) measurements are shown in Fig. 3.

**FIG.3.** Capacitance versus frequency (C-F) plot

The C-F plot shows the data representation in terms of capacitance instead of impedance. Fig. 3 shows the C-F plot of the three representative devices. From the C-F plot it is observed that there are two distinct capacitive plateaus for both devices, one at the 0.1 kHz range and another at 100 kHz range. The plateau at 100 kHz range is labeled Cbulk whereas the plateau at 0.1 kHz is labelled Cel. Cbulk is generally attributed to the capacitance arising from the various polarization effects of the perovskite photoactive layer (e.g. octahedra reorientation and ionic defects).[26] The second capacitive plateau, Cel was attributed to electronic and ionic accumulation to the interfaces of devices by previous reports.[26] As it was previously stated, Cel is predominantly affected by the photogenerated charge accumulation at the device interfaces. From the above figure it can be seen that Cel plateau is much lower for Cu:NiOx and even lower for CuO-based device compared to PEDOT:PSS-based device. This indicates decreased charge accumulation at the HTL/Pvsk interface and subsequently related to the increase in Voc and Jsc. The above observations are also in accordance with the increase in Rrec observed before.

In summary, we have shown that the HTL plays an important role to the behavior of a perovskite active layer formation and p-i-n solar cell device performance. By replacing the most commonly used organic PEDOT:PSS HTL with inorganic based Cu:NiOx and CuO HTLs, the PCE values of the corresponding solar cells increased from 8.44% to 11.45% and 15.7% respectively. The latter can be attributed to the better HTL/CH$_3$NH$_3$PbI$_3$ energy level alignment (better hole extraction) as well as improved crystallinity of the CH$_3$NH$_3$PbI$_3$ photoactive layer as shown by AFM studies. To further understand the mechanisms of hole extraction, impedance spectroscopy was used to study the solar cells under investigation. Through the ECM observations inhomogeneous capacitance distribution within the device area of perovskite solar cells using inorganic HTL could explain the need to replace the ideal Cdl capacitor with CPE. Despite their higher PCE values the solar cells using inorganic CuO and Cu:NiOx HTLs have inhomogeneous hole collection at the perovskite/inorganic interface. In contrast a homogeneous capacitor in the ECM can provide a good fit for the less efficient, but with uniformly distributed charge and more reliable perovskite solar cells using the organic PEDOT:PSS HTL. Impedance spectroscopy studies have also shown that the increase in Voc and Jsc for the perovskite solar cells using inorganic HTLs was a result of the overall decrease in charge accumulation of the devices with Cu:NiOx and especially CuO both at the grain boundaries area as well as the HTL/Pvsk interface. The reduced charge accumulation for inorganic HTL based sperovskite solar cells under study could potentially be attributed to the reduced grain boundaries area (as a result of the increased grain size) as well as to the increased Rrec. The decrease in charge accumulation at the HTL/Pvsk interface was also related to the reduction in the values of Cdl and Cel for the solar cells incorporating Cu:NiOx and CuO inorganic based HTLs. These results contribute towards the understanding of the charge accumulation and hole selectivity of p-i-n perovskite solar cells and can provide information relevant to HTL materials design/selection procedures for high performance perovskite based photovoltaics.

**Acknowledgements**

This project has received funding from the European Research Council (ERC) under the European Union's Horizon 2020 research and innovation programme (grant agreement No 647311).

[1] Kojima, A. et. al., *"Organometal Halide Perovskites as Visible-Light sensitizers for Photovoltaic Cells"*, J.Am.Chem.Soc, **131 (17)**, 6050, 2009.

[2] Dyesol LTD, New world record of PSC Efficiency, Media Release, (Ed: Dyesol Ltd), 2015.

[3] Choi, H. et.al, *"Conjugated polyelectrolyte hole transport layer for inverted-type perovskite solar cells"*, Nat. Comm., **6**, 2015.

[4] Snaith, H. et. al., *"Perovskites: The emergence of a new era for low cost, high efficiency solar cells"*, J.Phys.Chem.Lett., **4**, 3623-3630, 2013.

[5] Stoumpos, C.C. et.al, *"Semiconducting Tin and Lead Iodide Perovskites with organic cations: Phase Transitions, High Mobilities and near-infrared Photoluminescent Properites"*, Inorg.Chem, **52 (15)**, 9019-9038, 2013.

[6]

[7] Eperon, G.E. et.al, *"Formamidinium lead trihalide: a broadly tunable perovskite for efficient planar heterojunction solar cells"*, Energy Environ. Sci, **7 (3)**, 982-988, 2014.

[8] Liu, Y. et.al, *"Understanding Interface Engineering for High-Performance Fullerene/Perovskite Planar Heterojunction Solar Cells"*, Adv. Energ.Mat. 2015.

[9] Noh, J.H. et.al, *"Chemical management for colorful, efficient, and stable inorganic-organic hybrid nanostructured solar cells,* Nano Lett., **13**, 1764-1769, 2013.

[10] Wehrenfenning C. et.al, *"High Charge Carrier Mobilities and Lifetimes in Organolead Trihalide Perovskites"*, Adv. Mat, **26**, 1584-1589, 2014.

[11] Whang, Z.K. et.al, *"Induced Crystallization of Perovskites by a Perylene udnerlayer for High-Performance Solar Cells"*, ACS Nano, **10 (5)**, 5479-5489, 2016.

[12] Chen,J. et.al, *"Origin of the High Performance of perovskite solar cells with large grains"*, Appl.Phys.Lett, **108**, 2016.

[13] Zinab, H.B. et.al, *"Advances in Hole transport materials engineering for stable and efficient perovskite soalr cells"*, Nano Energy, **34**, 271-305, 2017.

[14] Calio L. et.al, *"Hole-Transport Materials for perovskite soalr cells"*, Angewandte Chemie, **55 (47)**, 14522-14545, 2016.

[15] Chin H.T. et.al, *"A review of organic small molecule-based hole-transporting materials for meso-structured organic-inorganic perovskite soalr cells"*, J.Mater.Chem. A, **4**, 15788-15822, 2016.

[16] Liang, P.W et.al, *"Roles of fullerene-based interlayers in enhancing the performance of organometal perovskite thin-film solar cells",* Advanced Energy Materials, 2015.

[17] Jung J. et.al, *"A low-temperature, solution-processable Cu-doped Nickel oxide hole-transporting layer via the combustion method for high-performance thin-film perovskite solar cells",* Advanced Materials, **27**, 2015.

[18] Savva A. et.al, *"Room Temperature Nanoparticulate Interfacial Layers for Perovskite Solar Cells via solvothermal synthesis",* Submitted to Journal of Material Chemistry A, 2017.

[19] Kwon J.D, et.al, *"Controlled growth and properties of p-type cuprous oxide films by plasma-enhanced atomic layer deposition at low-temperature"*, Applied Surface Science, 2013.

[20] Jung, J. et.al, *"A low-temperature solution processable, Cu-doped nickel oxide hole-transporting layer via the combustion method for high-performance thin-film perovskite solar cells",* Advanced Materials, **27 (47)**, 7874-7880, 2015.

[21] Li Y et.al, *"Highly efficient p-i-n perovskite solar cells utilizing novel low-temperature solution processed hole transport materials with linear π-conjugated structure"*, Small, **35**, 4902-4908, 2016

[22] Chueh C. et.al, *"Recent progress and perspective in solution-processed interfacial materials for efficient and stable polymer organometal perovskite soalr cells"*, Energy. Environ. Sci, 2015.

[23] Bi.C et.al, *"Non-wetting surface-driven high-aspect-ration crystalline grain growth for efficient hybrid perovskite solar cells"*, Nat.Comm, **6**, 7747, 2015.

[24] Bag, M. et. al., *"Kinetics of Ion transport in perovskite active layer and its implications for active layer stability"*, J.Am.Chem.Soc, **137**, 13130-13137, 2015.

[25] Liu, Y. et.al, *"Understanding interface engineering for high performance fullerene/perovskite planar heterojunction solar cells"*, Advanced Energy Materials, 2015.

[26] Guerrero, A. et.al, *"Properties of contact and bulk impedances in hybrid lead halide perovskite solar cells including inductive loop elements"*, J.Phys.Chem, 2016.

[27] More-Sero I. et.al, *"Implications of the negative capacitance observed at forward bias in nanocomposite and polycrystalline silicon solar cells"*, Nano Letters, **6 (4)**, 2006.[25] Radvanyi, E.

[28] Radvanyi E. et.al, *"Study and modelling of the solid electrolyte interphase behavior in nano-silicon anodes by electrical impedance spectroscopy"*, Electrochemica Acta, **137**, 751-757, 2014

[29] Yuan Y, et.al, *"Ion migration in organometal trihalide perovskite and its impact on photovoltaic efficiency and stability"*, Accounts of chemical research, **49 (2)**, 286-293, 2016.

[30] Savva A. et.al, *"Photovoltaic analysis of the effects of PEDOT:PSS-additives hole selective contacts on the efficiency and lifetime performance of inverted organic solar cells"*, Solar Energy Materials and Solar Cells, **132**, 507-514, 2015.

[31] Jundale D.M. et.al. ,*"Nanocrystalline CuO thin films, synthesis microstructural and optoelectronic proeprties",* J.Mater.Sci.Mater.Elctron. ,**23**, 1492-1499, 2012.

[32] Mulder, W.H. et.al, *"Tafel current at fractal electrodes: Connection with admittance spectra",* J. Electroanal.Chem, **285**, 1990.

[33] Joric, J.P et.al, *"CPE analysis by local electrochemical impedance spectroscopy",* Electrochemica Acta, **51**, 1473-1479, 2006.

[34] Schiller, C.A, *"The evaluation of experimental dielectric data of barrier coatings by emans of different models",* Electrochemica Acta, **46**, 2001.

[35] Vitoratos E. et.al, *"Thermal degradation mechanisms of PEDOT:PSS"*, Organic Electronics, **10**, 61-66, 2009.

# List of Figures

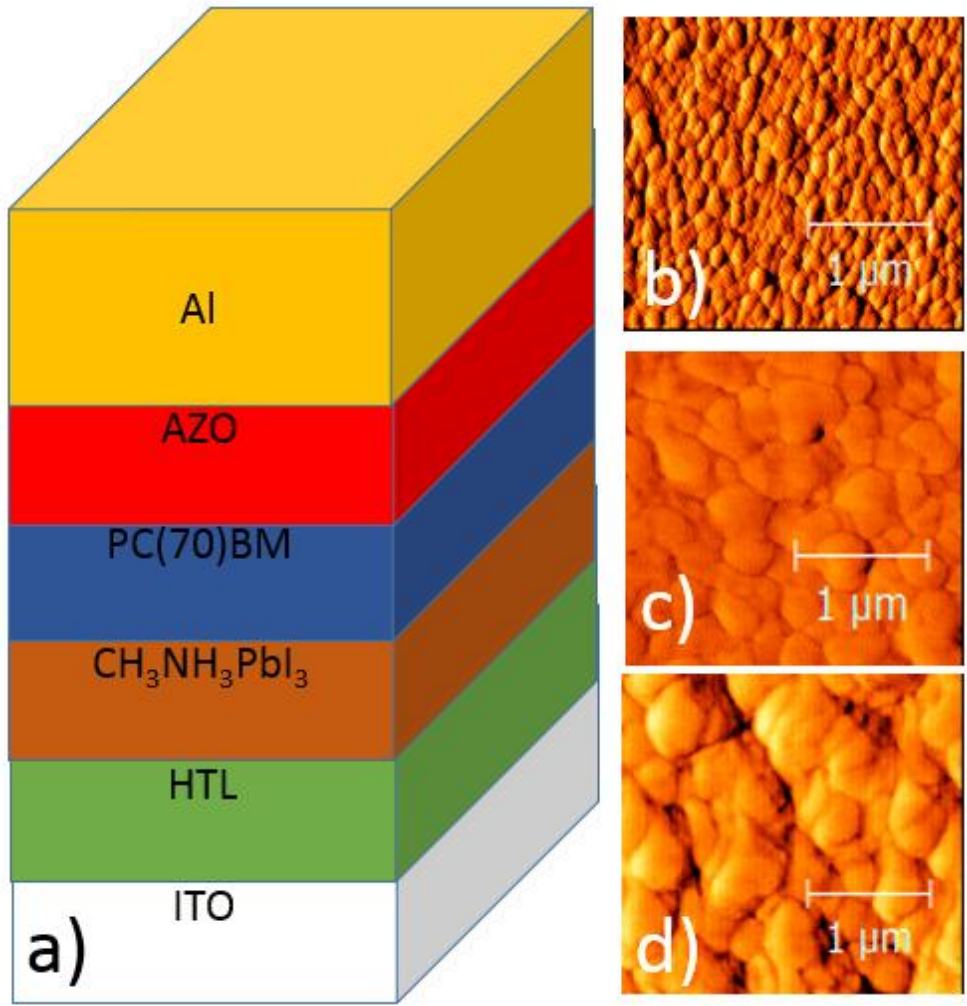

**FIG.1. a)** Graphical presentation of the solar cell layers, **b)** AFM measurements of ITO/PEDOT: PSS/CH$_3$NH$_3$PbI$_3$, **c)** ITO/Cu:NiOx/CH$_3$NH$_3$PbI$_3$ and **d)** ITO/CuO/CH$_3$NH$_3$PbI$_3$.

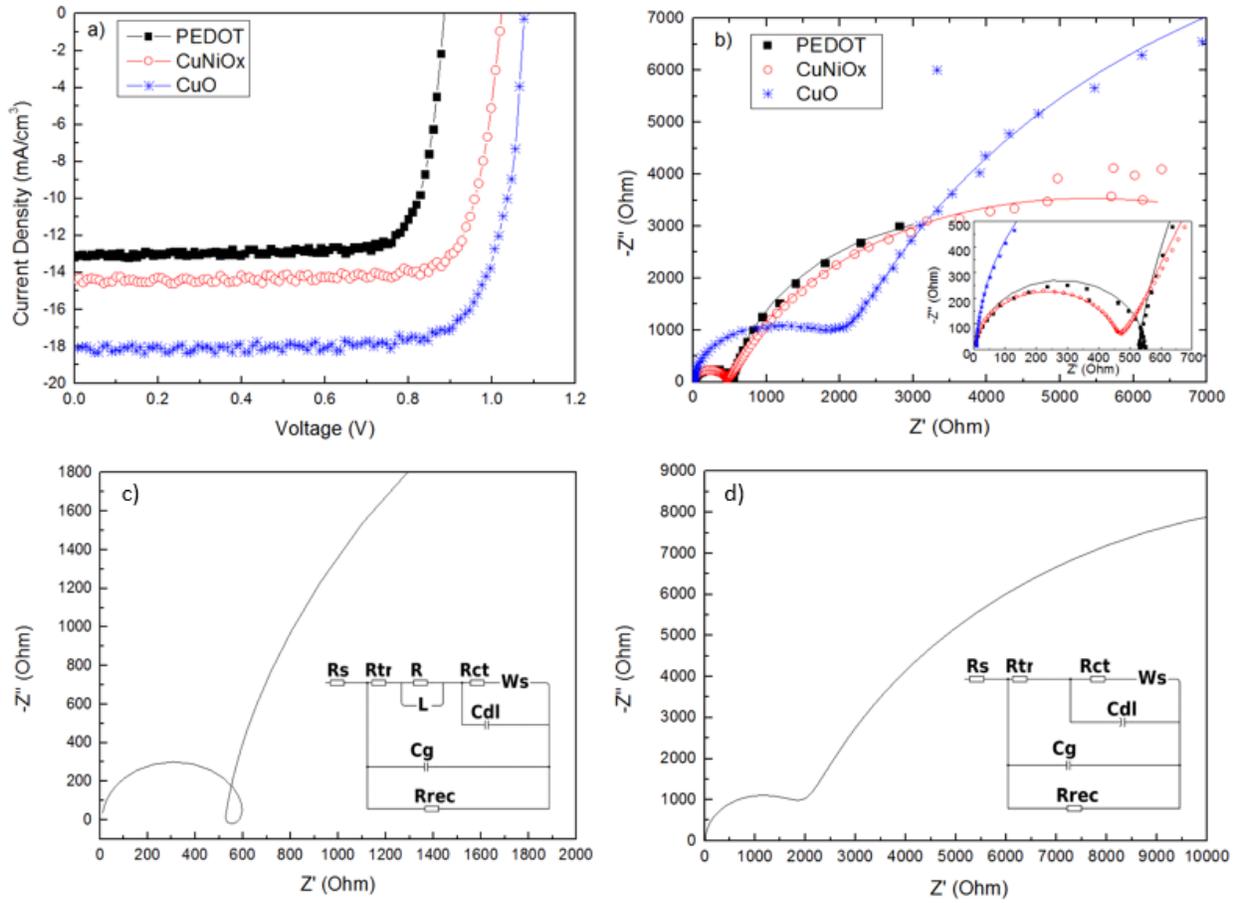

**FIG.2. a)** Current density versus voltage (J-V) characteristics under illumination, **b)** Nyquist plots **c)** Equivalent circuit model and simulation curve used for analyzing PEDOT: PSS based devices, **d)** Equivalent circuit model and simulation curve used for analyzing Cu:NiOx and CuO based devices.

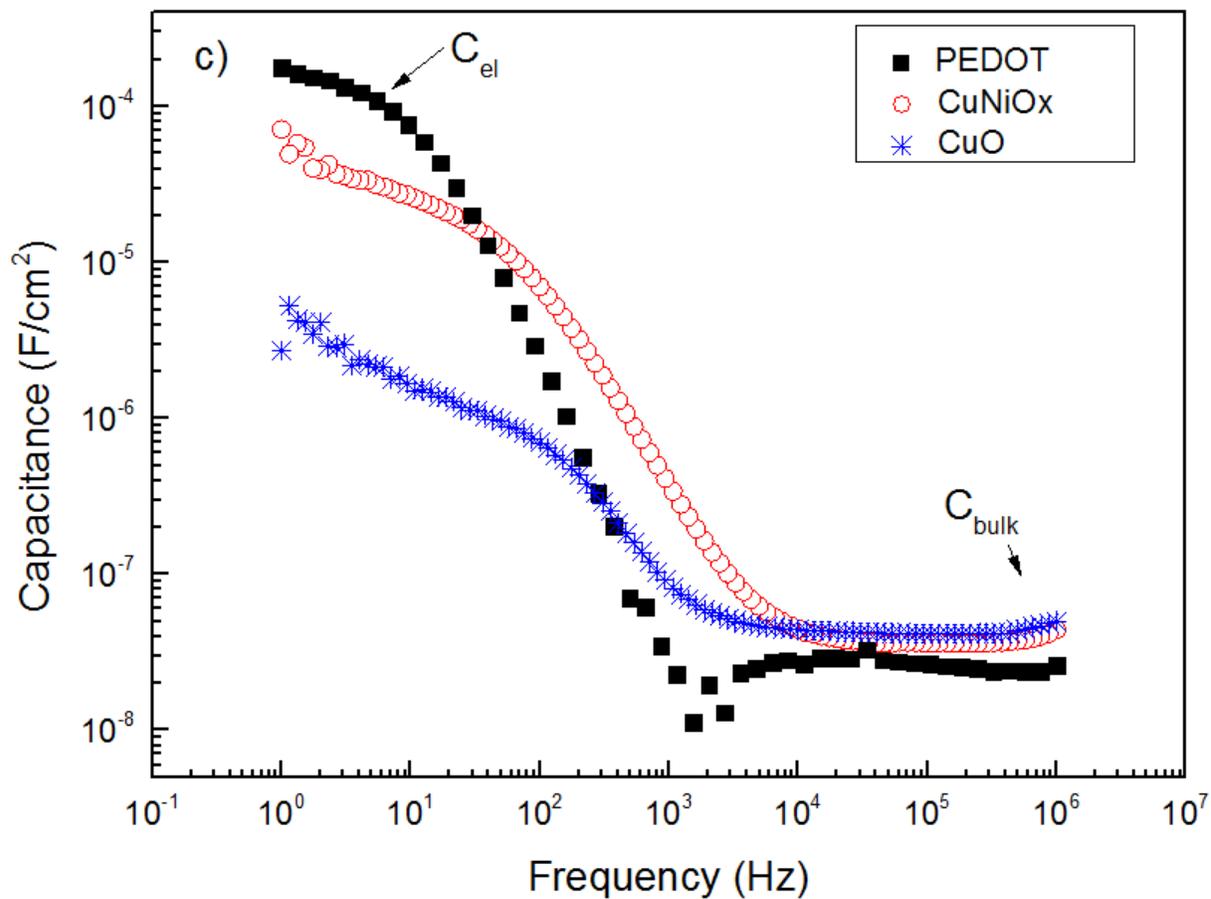

**FIG.3.** Capacitance versus frequency (C-F) plot